\documentstyle[preprint,prl,aps]{revtex}
\begin{document}
\draft
{\noindent \bf Comment on ``Energy Dependence of Electron Lifetime in Graphite 
Observed with Femtosecond Photoemission Spectroscopy''}

In a recent letter, Xu {\it et al.} \cite{sx1}
reported measurement of electron lifetime in graphite with photoemission 
spectroscopy. By performing a least-square fitting of the observed 
energy dependence of the electron lifetime and adapting  
a layered two-dimensional model for graphite, the
authors attributed the measured electron-electron scattering rate 
to contributions due to 
acoustic plasmon excitations. 
They came to the conclusion that for a
layered two-dimensional electron gas
with layer separation small compared with the
average inter-electron distance
within the layer, 
the acoustic plasmon excitations dominate electron-electron scattering 
processes and give rise to a linear 
energy dependence of the scattering rate 
for electrons with energies close to the Fermi energy.
The authors further implied that the 
scattering due to acoustic plasmon excitations 
may also be applied to explain
the linear energy and temperature dependence of 
carrier lifetimes observed 
in the normal state of high-$T_c$ superconductors
since most of the high-$T_c$ superconductor materials have 
similar layered electronic structures.
The purpose of this comment is to point out that these authors'
theoretical interpretations of their experimental data
and, most importantly, the suggested implication for 
high-$T_c$ superconductors 
are incorrect and misleading.
 
Electron-electron scattering contains contributions from both
single particle electron-hole excitations and collective 
plasmon excitations. 
In a layered electronic system, the plasmon spectrum consists of
optical and acoustic modes \cite{sd1}. 
The scattering rate due to the acoustic plasmon excitations 
can be expressed analytically in a narrow energy window as \cite{sx1,th1}
\begin{equation}
{1\over\tau_{ac}(k)}=C (\sqrt{\epsilon_k}-\sqrt{\epsilon_c})^2,
\label{equ:e1}
\end{equation}
where $C$ is a constant, 
$\epsilon_k=\hbar^2k^2/2m$ is electron energy,
and $\epsilon_c$ is a threshold energy.
The above expression is valid only for electrons with
energies $\epsilon_c <\epsilon_k<\epsilon^*$ \cite{th1}:
for electrons with  energies smaller than $\epsilon_c$, plasmon 
excitations do not contribute to electron scattering because of 
restrictions from
momentum and energy conservations; 
for electrons with energies higher than $\epsilon^*$,
$\tau_{ac}(k)^{-1}$ actually decreases as a function of $\epsilon_k$,  
and Eq.(\ref{equ:e1}) does not apply.
Since the threshold energy $\epsilon_c$ is higher 
than the Fermi energy and decreases with decreasing layer separation
\cite{th1},
the acoustic plasmons
do not contribute to scattering of electrons with 
energies close to the Fermi surface, 
except for extremely small layer separations where 
$\epsilon_c$ is sufficiently close to $\epsilon_F$.
In this small layer separation limit, 
Eq.(\ref{equ:e1}) gives 
$\tau_{ac}(k)^{-1}\sim(\epsilon_k-\epsilon_F)^2$ for
$\epsilon_k\rightarrow\epsilon_c\simeq\epsilon_F$, 
which is quadratic, not linear, in energy. 
We conclude, therefore, that in sharp contrast to statements made
in ref. \cite{sx1}, 
the electronic scattering rate due to acoustic plasmon excitations 
does not show linear energy dependence under any circumstances.
The theoretical conclusions of ref. \cite{sx1} were obtained 
by comparing the measured data with Eq.(\ref{equ:e1}),
which, as should be obvious from our comment, does not show a linear
$(\epsilon-\epsilon_F)$ behavior for any values of electron energy.
An additional problem is 
that the experimental data, with electron energies 
on the order of $eV$,  
are in the high electron energy regime 
where Eq.(\ref{equ:e1}) is not valid.
It is then clear that the theoretical interpretations of 
ref. \cite{sx1} are wrong.

There is another problem associated with attributing the measured 
scattering rate solely to
contributions due to acoustic plasmon excitations.
In a layered electronic gas,
the spectral weight of acoustic plasmons
decreases as the layer separation decreases,
since only the optical plasmon mode survives in the zero
layer separation limit.
In graphite, the layer separation is $3.7\AA$ and 
the average inter-electron distance within the layer
is on the order of $100\sim200\AA$.
Under this small layer separation condition,
contributions to the scattering rate from excitations 
other than the acoustic plasmons may not necessarily be negligible.
The authors of ref. \cite{sx1} uncritically disregarded 
other possible sources of 
contributions without any quantitative examination.

We emphasize that the implication of ref. \cite{sx1}
for carrier lifetimes in the normal state
of high-$T_c$ superconductors, as suggested by the 
authors, is invalid.  
Many experiments on the normal state of 
high-$T_c$ superconductor materials 
suggest linear energy and temperature dependence of lifetimes
for carriers close to the Fermi energy.  
It is well known that the anomalous linear 
energy and temperature dependence of
the lifetime is not expected from a 
Fermi-liquid type many-body theory for a single-layer uniform
two-dimensional electron gas.  
In fact, the standard Fermi-liquid result for the scattering rate of
a two-dimensional electron gas is an 
$(\epsilon-\epsilon_F)^2\ln(\epsilon-\epsilon_F)$ behavior
for $\epsilon$ close to $\epsilon_F$, which is extremely 
well-verified experimentally \cite{me1}.
From the discussion above, it is clear that acoustic plasmon
contributions in a layered
two-dimensional electron gas do not
produce a linear $(\epsilon-\epsilon_F)$ behavior 
for the scattering rate either.
It is well-established \cite{sha1} 
that both two- and three-dimensional electron 
systems, at least within the standard perturbative many-body
picture, have scattering rates which 
are essentially quadratic in energy, {\it i.e.}
$(\epsilon-\epsilon_F)^2$ within logarithmic corrections, 
for electrons close to the Fermi surface.
The same is also true for a layered electron gas system,
at least within the standard Fermi liquid many-body perturbation theory.
\vskip 1.5cm
\leftline{Lian Zheng and S. Das Sarma}
\leftline{\ \ \ Department of Physics}
\leftline{\ \ \ University of Maryland}
\leftline{\ \ \ College Park, Maryland 20742-2111}
\vskip 1cm
\leftline{PACS numbers: 73.20Dx, 73.20.Mf, 73.61.-r}

\end{document}